\documentclass[Afour,sagev,times]{sagej}

\usepackage{moreverb,url}
\usepackage{enumitem}
\usepackage[perpage]{footmisc}
\usepackage{xcolor}

\usepackage[colorlinks,bookmarksopen,bookmarksnumbered,citecolor=red,urlcolor=red]{hyperref}

\newcommand\BibTeX{{\rmfamily B\kern-.05em \textsc{i\kern-.025em b}\kern-.08em
T\kern-.1667em\lower.7ex\hbox{E}\kern-.125emX}}

\begin{document}

\runninghead{Gerardi et al.}

\title{Active Informed Consent to Boost the Application of Machine Learning in Medicine}

\author{Marco Gerardi\affilnum{1}, Katarzyna Barud\affilnum{2}, Marie-Catherine Wagner\affilnum{2}, 
Nikolaus Forgò\affilnum{2}, 
Francesca Fallucchi\affilnum{1}, Noemi Scarpato\affilnum{3},  Fiorella Guadagni\affilnum{3,4} and  Fabio Massimo Zanzotto\affilnum{5}}

\affiliation{\affilnum{1}Guglielmo Marconi University, Rome, Italy, 
\affilnum{2}University of Vienna, Vienna, Austria, 
\affilnum{3}San Raffaele Roma Open University, Rome, Italy,
\affilnum{4}IRCCS San Raffaele Roma, San Raffaele Roma Open University,
\affilnum{5}University of Rome Tor Vergata, Rome, Italy}

\corrauth{Fabio Massimo Zanzotto,
University of Rome Tor Vergata,
Viale del Politecnico, 1,
00133, Rome, Italy.}

\email{fabio.massimo.zanzotto@uniroma2.it}

\begin{abstract}
Machine Learning may push research in precision medicine to unprecedented heights.
To succeed, machine learning needs a large amount of data, often including personal data.
Therefore, machine learning applied to precision medicine is on a cliff edge: if it does not learn to fly, it will deeply fall down.
In this paper, we present Active Informed Consent (AIC) as a novel hybrid legal-technological tool to foster the gathering of a large amount of data for machine learning. We carefully analyzed the compliance of this technological tool to the legal intricacies protecting the privacy of European Citizens.
\end{abstract}


\maketitle

\section{Introduction}

Machine Learning has the potential to revolutionize research in precision medicine, as early results are suggesting. Intelligent machines may help in discovering new drugs  \cite{zhang_machine_2017:drug_discovery,vamathevan_applications_2019:drug_discovery,dara_machine_2022:drug_discovery}, may contribute in analyzing medical images to spot nuances that 
 clinical doctors may overlook \cite{maier_gentle_2019:image_analysis,zhang_radiological_2019:image_analysis}, may reduce the burden of generating repetitive medical reports \cite{babar_evaluating_2021:image_analysis}, and may also help in predicting clinical outcomes \cite{chen_machine_2017:prediction}. All these areas  may be positively affected by the use of these technologies. These initial results are really promising.

To succeed in its promises in health research projects, machine learning needs a large amount of clinical and, eventually, omics data to infer regularities for proposing tailored therapeutic plans. In principle, data are available. Hospitals and national health systems are starting to accumulate data in databases or machine-readable files. Electronic health record (EHR) repositories store large sets of clinical and omics data. Moreover, subsets of these data may be made available for research purposes for specific research projects. This seems to be an ideal situation, but it hides an inherent limitation.

Machine learning (ML) applied to precision medicine is on a cliff edge: if it does not learn to fly, it will deeply fall down. Indeed, ML needs data, and patients in most cases donate their data\cite{Zanzotto2019243} for processing to specific research projects by signing informed consent. 
Hence, these data cannot be easily shared and gathered in larger datasets, which may make the difference for machine learning. To overcome this and allow research participants to agree to participate in further research projects, a dynamic consent approach has been broadly discussed in the scientific community. \cite{Budin-Ljosne2017, Teare2021, Kaye2015, pmid31729900}

In this paper, we present the Active Informed Consent (AIC) as a novel hybrid legal-technological tool to foster the gathering of a large amount of data for machine learning. AIC builds upon what dynamic consent \cite{budin-ljosne_dynamic_2017} proposes. 
Besides being a classical informed consent, AIC tool will help patients to donate data in a cascade manner. Indeed, AIC tool will be an application that will ask patients whether donated data can be transferred to other research projects. With AIC tool, patients or their legal representatives will become active actors in controlling for what purposes their personal data will be used. The solution will enhance their right to make autonomous choices regarding their participation in research and processing of their health data.

The paper is organized into two blocks: a legal block outlining legal requirements  and a technological answer satisfying these constraints. The compliance of the technological answer to the legal issues is then analyzed. Finally, we draw some conclusions in the last section.

\section{Consent in Health Research – an International Ethical and Legal Requirement?}
\label{sec:legal}

When processing personal data, in particular data concerning health and genetic data with new technologies such as machine learning, it is of most importance to carefully analyze and implement the requirements set out by ethical standards and the law. One should note that there is not one standard, which may be followed. Depending on the nature of the data and purpose, there are various ethical and legal requirements, which must be taken into consideration. Some jurisdictions may require to seek for consent to process data for medical research. At the same time, there is an ethical requirement stemming from The Declaration of Helsinki \footnote{World Medical Association Declaration of Helsinki - Ethical Principles for Medical Research Involving Human Subjects, adopted by the 18th WMA General Assembly, Helsinki, Finland, June 1964
and last time amended by the 64th WMA General Assembly, Fortaleza, Brazil, October 2013, https://www.wma.net/policies-post/wma-declaration-of-helsinki-ethical-principles-for-medical-research-involving-human-subjects/ }, an internationally recognized global standard for instance, which requires generally a consent from the patient for medical research involving human subjects. This means that in some instances, there may be some overlapping conditions set forth by the law or by an ethical standard, such as consent. Nevertheless, in order to avoid any confusion, it is important to keep these issues clearly separated. Also the EDPB (European Data Protection Board) stressed in its Guidelines on Consent \footnote{EDPB Guideline on Consent, 154.} that “[…] consent for the use of personal data should be distinguished from other consent requirements that serve as an ethical standard […]”. 

Ethical Requirements in contrast to obligations stemming from the law are usually not legally binding. However, the standards shall be followed, as sanctions and loss of reputation within an issuing association may be the consequence. Although consent is often also an essential ethical requirement enhancing patient autonomy and self-determination, this article focuses mainly on legal requirements, as ethical standards.

\subsection{Consent in the EU, US and Chinese Data Protection Legal Framework}
The consent as a legal and ethical requirement for processing of personal data is present in the European, but also non-European legislation. The EU, the US and Chinese legislation play a significant role in health research and where the topic of data subjects (i.e. research participant's) consent is of relevance for the successful conduct of research activities. 

In the European data protection legislation, the GDPR lays down the main principles and rules of processing personal data and therefore builds the fundament of all processing activities. The first principle, defined in Article 5(1)(a) GDPR, the principle of “lawfulness”, sets forth that all processing activities must be justified with a legal basis. This means that one of six possible legal bases laid down in Article 6 GDPR must be fulfilled. When processing sensitive data, such as health or genetic data, a controller must be particularly cautious. The GDPR generally prohibits the processing of such data unless one of the ten conditions outlined in Article 9(2) GDPR is fulfilled. Both Article 6 and Article 9 GDPR provide as first possibility consent and explicit consent respectively. This does not mean that consent is an absolute requirement to process sensitive data. It is one of the possible legal bases provided by the GDPR. Notwithstanding that, there are some European jurisdictions, where consent in some cases is an essential requirement for processing health or genetic data. This is, for example, a case of Irish legislation, which requires explicit consent under the GDPR as a requirement for processing of the data for primary and secondary health research.\cite{7a6a23e4a9864a0e80f47484b1bfe2cf}, \footnote{Regulation 3(1)(e) of Health Research Regulations 2018 states that explicit consent is one of the mandatory "suitable and specific measures" that must be in place for the processing of data for health research purposes unless the researcher has been granted a consent declaration under Regulation 5 or under the transitional arrangements (Regulation 6), see further: https://www.hrb.ie/funding/gdpr-guidance-for-researchers/gdpr-and-health-research/consent/explicit-consent/}  
This not unified approach within the EU is the result of various opening clauses proposed by the GDPR. 

Additionally, in the context of machine learning tools a controller must also take into consideration, when assessing the potential legal basis, Article 22(4) GDPR, which does not allow automated individual decision-making, including profiling, based on sensitive data, unless it is based on the explicit consent of a data subject (Article 9(2)(a) GDPR) or such processing is necessary  for reasons of substantial public interest, on the basis of Union or Member State law (Article 9(2)(g) GDPR). When choosing the legal basis, a controller should therefore assess whether the machine learning tool does make the decision automatically and whether Article 22(4) GDPR is applicable.

Using consent as the primary legal basis for processing personal data does not only have the above-mentioned advantages. As consent is also recognized as a legal basis in other countries that participate in the innovation race, this proposed solution has the potential to be used internationally. In the US and China, the development in health research achieves a very advanced level and the collaboration between Europe and these markets is very beneficial for scientific initiatives. Nevertheless, it is clear that these regimes differ, in some instances, fundamentally, especially in the privacy and data protection sphere. Throughout the years, we could observe the "Brussels effect", requiring the jurisdictions from outside of EU to reconsider the legal framework set for data protection. \footnote{Anu Bradford, 'The Brussels Effect' (2012) 107 Nw U L Rev 1}. 
It goes without saying that the EU has a stricter privacy regime than the US.\footnote{Anu Bradford, 'The Brussels Effect' (2012) 107 Nw U L Rev 1; Mark F. Kightlinger, Twilight of the Idols? EU Internet Privacy and the Post Enlightenment
Paradigm, 14 COLUM. J. EUR. L. 1, 2-3 (2007)} 
The EU has a comprehensive legislation establishing privacy principles not only for public, but also for private sector, whereas the US data privacy laws have been so far restricted mostly to the public sector and some sensitive ones, including health care and banking. Moreover, the EU has significantly different approach than China, whose digital authoritarianism has been broadly commented.\footnote{see for e.g. Lilkov D., Made in China: Tackling Digital Authoritarianism, May 20, 2020, accessible here: https://journals.sagepub.com/doi/full/10.1177/1781685820920121;}.

Nevertheless, all regimes recognize consent as a possible legal basis for processing sensitive data in health research context. In this publication, we will focus on these countries and their approach to consent, as this plays a significant role for AIC and the possibility to use this solution not only on the European, but also US and Chinese market.

In China, the data protection regime has been recently in a phase of changes. On 1 November 2021, the Personal Information Protection Law (further “PIPL”)\footnote{The text of the PIPL in Chinese and in English is available here: https://digichina.stanford.edu/work/translation-personal-information-protection-law-of-the-peoples-republic-of-china-effective-nov-1-2021/} entered into force, being the first such comprehensive data protection act in this country. It applies to the entities and individuals who carry out personal information processing activities within China and outside of China – if the purpose of processing is to offer goods or services to individuals in China or to monitor and evaluate the activities of individuals in China. The mechanism of personal information protection defined in the PIPL sets forth what is personal information and its sensitive type. It also introduces principles of personal information processing, as well as consent and non-consent bases for processing. According to the PIPL, personal information is any information relating to an identified or identifiable natural person. Personal information on biometric characteristics and on medical health represent sensitive data (Article 28 PIPL). Health data is not specifically defined under Chinese law. Article 13 sets out the legal bases for personal information processing. The primary one is consent (Article 13(1) PIPL). It should be noted that, unlike the GDPR, “legitimate interest” has not been included as a basis for data processing.  The individual consent shall be voluntary, explicit, and fully informed. (Article 14 PIPL). Moreover, whenever the purposes, means or category of processed information changes, a new consent shall be obtained from the individual. Every individual shall have the right to withdraw consent. The entity collecting and processing the data cannot disclose it, except where separate consent has been obtained from the individuals (Article 25 PIPL). 
According to Article 28 PIPL, sensitive data, such as medical health data, can be handled only where there is a specific purpose and a need to fulfill, and under circumstances of strict protection measures. Additionally, following the Article 29 PIPL, to handle sensitive personal information, the individual's separate consent shall be obtained. Moreover, where laws or administrative regulations provide that written consent shall be obtained for handling sensitive personal information, those provisions are to be followed.  
However, it should be emphasized that although Chapter II introduces some limitations for State authorities, e.g. by imposing the requirement to notify individuals about the personal information processing and to obtain their consent before handling their personal information, it also provides vague exemptions from these requirements, "where laws or administrative regulations provide that secrecy shall be preserved" or "where notification will impede State organs’ fulfillment of their statutory duties and responsibilities." (Article 35 PIPL).
Regardless the last point, it is clear that consent is one of the basis recognised by the Chinese law in the context of processing data related to health and the AIC tool could be of help for obtaining such consent in the context of health research, however, not necessarily when it comes to the processing of health data by State organs.  

In the United States, the requirement to obtain consent for processing special categories of data is more complex. There is no federal law regulating this matter, and specific state laws are introduced in this context. Nevertheless, the laws generally require that data subjects should be informed about data processing pre-collection, for example in a privacy policy. It should describe the collection, use, and disclosure practices, choices that data subjects have regarding their personal information as well as the contact information of the entity processing the data.
Opt-in consent is required to collect, use and disclose selected types of sensitive data, such as health information and biometric data. The HIPAA Privacy Rule\footnote{Health Insurance Portability and Accountability Act (HIPAA) lays down data protection and data confidentiality requirements applying to the processing of protected health information (PHI)}  introduces requirements which must be met before using or disclosing identifiable protected health information (“PHI”) for research\footnote{\url{https://www.hhs.gov/hipaa/for-professionals/special-topics/research/index.html}}, defined in this context as “a systematic investigation, including research development, testing, and evaluation, designed to develop or contribute to generalizable knowledge”\footnote{See HIPPA, 45 CFR 164.501}. This includes the development of research repositories and databases for research.\footnote{\url{https://privacyruleandresearch.nih.gov/pr_04.asp}}  
The health information can be used or disclosed when it has been de-identified\footnote{\url{https://privacyruleandresearch.nih.gov/pr_08.asp##8a}}. 
In such cases, the Privacy Rule foresees the possibility for waiver or alteration of the Privacy Rule Authorization requirement.  
If the research requires the processing of identifiable health information, additional requirements have to be met by so-called "covered entities" (a health plan, a health care clearinghouse, or a health care provider who transmits health information in electronic form in connection with a transaction for which HHS has adopted a standard) to provide researchers access to and use of PHI in their research activities. This includes obtaining authorization in accordance with HIPPAA Privacy Rule.
A valid Privacy Rule Authorization, to some extent analogue with the "explicit consent" defined in Article 9(2)(a) GDPR, is a permission (i.e. consent) signed by an individual, allowing a covered entity to use or disclose an individual’s PHI for the purposes and to the specified recipient(s). When this Authorization is obtained for research purposes, it is required that it pertains to a specific research study. It cannot refer to nonspecific or future, unspecified projects. This authorization, similar to the GDPR consent, differs from informed consent in the ethical understanding, as it focuses on privacy aspects and defines how, why and to whom the health information will be used for research. It should include the core elements and required statements defined in the Privacy Rule.\footnote{\url{https://privacyruleandresearch.nih.gov/pr_08.asp##8a}}  An alternative to obtaining an Authorization could be, in the case of use or disclosure of a limited data set including PHI, signing a data use agreement with the researching entity. Such a data set cannot include specific direct identifiers, as defined in the HIPPA Privacy Rule. 

Taking into consideration the above-mentioned legislation, there is no doubt that consent of the data subjects may play a significant role in the context of scientific research involving processing of their data concerning health. The AIC tool proposed in this paper may allow for appropriate collection of consent and its update, when requested by the data subject and where necessary in the research study, not only in the EU, but possibly also on non-EU markets considered above. 

\subsection{Key Elements of the Consent from a European Perspective} 
As the EU has one of the strictest data protection laws, this subsection as well as the Table \ref{tab_EC} provide an overview of the elements of the consent required by the GDPR.\footnote{Article 4(11), Article 7 and Recitals 32, 33, 42 and 43 GDPR.}  The consent must be freely given, specific, informed and unambiguous, and in case of processing special categories of data, which clinical and omics data represent, it has to be also explicit. 

\begin{table}
\begin{tabular}{|l|p{5cm}|}  
\hline  
Consent Element & Short Description \\
\hline  
Explicit& Present options to agree or disagree and make 
sure the data subject clearly indicates this/her
choice. \\ 
\hline
Freely Given & Give a real choice and the data subject shall not
feel compelled to consent  \\ 
\hline
Specific & Specify the purpose of each processing operation
in a transparent, clear, and understandable way \\ 
\hline
Informed &\vspace{-5mm}
\begin{enumerate}[label=\alph*.]
    \item the controller's identity,\vspace{-3mm}
    \item the purpose of each of the processing operations for which consent is sought,\vspace{-3mm}
    \item what (type of) data will be collected and used,\vspace{-3mm}
    \item the existence of the right to withdraw consent,\vspace{-3mm}
    \item information about the use of the data for automated decision-making in accordance with Article 22 (2) (c) where relevant, and \vspace{-3mm}
    \item on the possible risks of data transfers due to the absence of an adequate decision and of appropriate safeguards as described in Article 46.
\end{enumerate}\\
\hline
Unambiguous & e.g. use boxes to be ticked by the data subject.
Pre-ticked boxes are not allowed \\
\hline
\end{tabular}  
\caption{Elements of consent}
\label{tab_EC}

\end{table}

The data subject must have a real choice and not feel compelled to consent.\footnote{European Data Protection Board, 'Guidelines 05/2020 on consent under Regulation 2016/679’ (2020) 7.}  
According to the EDPB, the controller must actively inform about certain elements to enable a free choice of the data subject\footnote{European Data Protection Board, 'Guidelines 05/2020 on consent under Regulation 2016/679’ (2020) 15.}:

a.	the controller’s identity,
b.	the purpose of each of the processing operations for which consent is sought,
c.	what (type of) data will be collected and used, 
d.	the existence of the right to withdraw consent,
e.	information about the use of the data for automated decision-making in accordance with Article 22 (2)(c) where relevant, and
f.	on the possible risks of data transfers due to the absence of an adequate decision and of appropriate safeguards as described in Article 46.

The above-mentioned elements are crucial to be able to seek valid consent. One should note that the information provided in the consent form does not lift the additional controller’s obligation laid down in Articles 13 and 14. \footnote{European Data Protection Board, 'Guidelines 05/2020 on consent under Regulation 2016/679’ (2020) 15.} However, there are some overlaps between these two obligations.

Generally, it is strongly recommended to seek written consent, although orally given consent can be valid too. In a digital context, the controller may also consider seeking consent in an electronic form. Such as filling an electronic form or sending an email or uploading a document with the signature of the data subject.\footnote{European Data Protection Board, 'Guidelines 05/2020 on consent under Regulation 2016/679’ (2020) 21.}  A controller should consider layering the information provided to the data subject in order to ease the navigation through all information.\footnote{European Data Protection Board, 'Guidelines 05/2020 on consent under Regulation 2016/679’ (2020) 16.}  A controller may also consider the usage of standardized icons to enhance the transparency principle.\footnote{Article 12(7) GDPR.}  
In order to comply with the requirement of unambiguity, the controller could use boxes to be ticked by the data subject. Pre-ticked boxes are not allowed.\footnote{European Data Protection Board, 'Guidelines 05/2020 on consent under Regulation 2016/679’ (2020) 18.}  
If those requirements are not met, this legal basis may be considered invalid, and if there is no other legal basis applicable\footnote{This interpretation is supported by the European Commission’s GDPR Q\&A website, available at \url{https://ec.europa.eu/info/law/law-topic/data-protection/reform/rules-business-and-organisations/principles-gdpr/purpose-data-processing/can-we-use-data-another-purpose_en}} – the processing activity may be considered unlawful and infringing Article 6 and 9 GDPR, which may lead to severe financial liability (Article 83(5)(a) GDPR).

In order to seek valid consent, it is also crucial to specify the purpose of each processing operation (“specific consent”). The purpose is used to indicate the "why" it is necessary to collect the data of the interested party, indicating explicit and well-defined legitimate reasons. In accordance with Article 5(1)(b) GDPR, it is necessary to explain the motivation, or rather the purpose for which the data is used, in a transparent, clear, and understandable way as well as with exhaustive documentation. When personal data in a large scale is processed for scientific research, it may become difficult to set a specific purpose before the research starts. For this reason, recital 33 GDPR \footnote{Recitals to EU laws are not in themselves legally binding in the same way that the operative provisions are. However, where an EU law is ambiguous, the recitals are important in interpreting the ambiguous provision. It may be taken into consideration by the courts in the course of its interpretation to EU law, in the case of ambiguity in a particular provision within the legislation.} recognizes the possibility to formulate the consent for research in a broader manner:

\textit{(…) is often not possible to fully identify the purpose of personal data processing for scientific research purposes at the time of data collection. Therefore, data subjects should be allowed to give their consent to certain areas of scientific research when in keeping with recognized ethical standards for scientific research. }\footnote{Recital 33 GDPR}

Nevertheless, the EDPB stresses that recital 33 should not be interpreted as allowing a controller to circumvent the key principle of specifying purposes for which consent is asked.\footnote{European Data Protection Board, ‘Document on response to the request from the European Commission for clarifications on the consistent application of the GDPR, focusing on health research (2021) 26}  
If research purposes cannot be specified precisely, a controller must consider other ways to ensure that the requirements for consent are fulfilled best. The EDPB states that a controller can allow data subjects to consent for specific stages of a research project that are already known to take place or to describe purposes in a more general level. Moreover,  with regard to health data, the elastic approach from recital 33 is subject to a stricter interpretation due to the sensitivity of the data, and an advanced level of scrutiny\footnote{ EDPB Guidelines 05/2020 on consent under Regulation 2016/679, 157.}. It should be noted that this recital does not take precedence over the requirements for consent defined directly in the core text of the GDPR.\footnote{Requirements for consent defined in Articles 4(11), 6(1)(a), 7 and 9(2)(a) GDPR.} Taking those into consideration, the controller must seek for a new consent if further processing activities are outside of the scope of the purpose for which the personal data have been collected.



\section{The Active Informed Consent Platform}
\label{sec:system}

This section describes the idea proposed to solve the age-old issue of informed consent to be submitted and signed by the data subject.
Firstly, however, it is necessary to describe the two main technologies we will use: the IPFS distributed file system and the Ethereum blockchain.
Secondly, we will describe our solution.

\subsection{Technology Background}
\subsubsection{InterPlanetary File System:}
InterPlanetary File System (IPFS)\footnote{\url{https://ipfs.io/}} was born from the idea and effort of Juan Benet in 2014 with the goal of creating a very fast system for managing scientific data and, at the same time, keeping track of the changes made to them (versioning).
The core of IPFS is composed of a file management system with integrated versioning management that thus allows keeping track, over time, of any changes made to the files. 
From a technical point of view, IPFS is a synthesis of well-tested and widely used web technologies such as DTH (Distributed Hash Tables), BitTorrent technology, the Git versioning system, and the SFS (Self-certified File System) technique which, working in synergy with each other, allow the creation of a Peer-to-Peer swarm for the exchange of IPFS objects. 

The totality of IPFS objects forms a cryptographically authenticated data structure known as a Merkle DAG, which in turn may be used to model many other data structures. 
In this network there are no privileged nodes, each node keeps track of a portion of IPFS objects and exchanges them with other nodes in the network. 
The declared goal of IPFS, as stated in his white paper\footnote{\url{https://ipfs.io/ipfs/QmV9tSDx9UiPeWExXEeH6aoDvmihvx6jD5eLb4jbTaKGps}}, is to replace the HTTP protocol and build a better and more efficient web system for everyone, with the following strengths\cite{bcipfs, 9045940}:
\begin{itemize}
\item The ability to distribute large amounts of data with high efficiency, saving the used space since duplicates are not allowed.
\item The maintenance of each version of the archived files. 
\item Decentralizing the web and preventing centralized control of the contents.
\item Supporting the creation of resilient networks that can provide permanent availability even without a connection to the Internet backbone.
\end{itemize}

\begin{figure}[ht]
\includegraphics[width=10 cm]{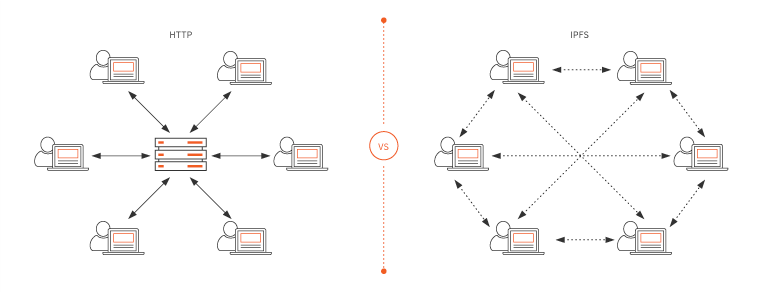}
\caption{Difference between the HTTP and IPFS protocols.\label{fig1}}
\end{figure}

When a file is uploaded to an IPFS node, it is identified with a unique fingerprint called a cryptographic hash and divided into blocks that are also marked with a fingerprint.
The IPFS infrastructure is designed to remove any duplicates from the network. Moreover, any change to any file, even if minimal, involves the variation of the associated hash code and therefore makes the file different from the original one, thus identifying any alterations. 
Each node in the network will store only the information of its interest and will maintain an index containing useful information about other objects available in the network.
In case of a file search in the network, the network returns the list of nodes that have stored the data related to the unique hash code of the object that is being searched, then subsequently, these nodes will be queried to download the requested object.
In addition, IPFS provides an information retrieval service similar to DNS to make it easy to store the names of objects stored on the network.

\subsubsection{Ethereum Blockchain:}

The blockchain can be seen as a particular distributed ledger with a distinct set of features and operational processes. 
Specifically, blockchains group transactions into blocks by protecting them against tampering with powerful cryptographic systems, each block is sequentially linked to all other blocks in the chain. 
These blocks, once inserted into the chain, are immutable and can be used to reconstruct the history of all the data within the blockchain.
In addition, there is a consensus system (e.g. Proof of Work - POW or Proof of Stake - POS) to determine how new blocks should be added to the blockchain.
The most famous use of the blockchain is the technology behind bitcoin.
In this case, the blockchain records currency transactions, i.e. the transfer of currency from one wallet to another.

This is just one of the possible uses of the technology, blockchains are useful in several other industrial fields and for many other types of records that go far beyond financial transactions. 
They can be used for all information that needs to be recorded in an immutable way such as health data in medical records\cite{bcipfs}, contracts, property transfers, purchase of goods and services, and much more.

Blockchain technology has the following features\cite{1032154406,1152290190,bttech}:
\begin{itemize}

\item \emph{Decentralized} \cite{8287055}: This is perhaps the most important concept of a blockchain, as there is no need for a third party or intermediary to vouch for the validation of transactions and information entered into the distributed database. A consensus mechanism shared and accepted by all nodes in the network is used to validate transactions and blocks. For this reason, the Blockchain is defined as a trustless system. 

\item \emph{Distributed}\cite{10.1093/jamia/ocx068}: Because the data is replicated across all nodes in the network, the system is highly resistant to failure or cyber-attacks.
Since each node has a copy of the entire database, there is no single point of failure, and any damage to a peer in the network does not affect the overall operation of the system, unlike in centralized systems where a cyber-attack or a blackout of the data center can interrupt the service of an entire system. This redundancy gives the entire network high availability and reliability.

\item \emph{Transparent}\cite{art/transp}: The data contained in the blockchain is available to all nodes in the network and, in the case of public networks, it is also freely available to anyone accessing the network. This feature provides the blockchain with the peculiarity of being transparent, and therefore trust in the system is also guaranteed by this feature.

\item \emph{Immutable}\cite{Yaqoob2022}: Once the data has been written in the blockchain, it is extremely difficult to modify it, as this would involve a huge expenditure of computing power. If, for example, you wanted to modify a transaction that was written 10 blocks ago, a hacker would have to modify and recalculate all 10 blocks or cut the chain and invalidate the next 9 blocks.

\item \emph{Highly secure}: All transactions on the blockchain are encrypted, so it gives the network high integrity.

\end{itemize}

For our project, we will use the blockchain of the Ethereum ecosystem, as it enables the use of smart contracts \cite{Szabo_1997,buterin2009next}.
A smart contract can be defined as an event-driven program, written in a specific programming language (depending on the blockchain on which it is deployed), executed on the nodes of a blockchain that results in a change of state of the distributed ledger.
The state change is generally represented by a transaction that changes the blockchain.
When a smart contract is invoked, the expected outcome is purely deterministic based on data provided as input.
Since one of the main characteristics of the blockchain is its quality of acting as a decentralized database shared between all actors that are part of the network, without the involvement of intermediaries or third parties, this peculiarity allows the smart contract to be executed without the possibility of alteration, in a certain way, saving time and money, reducing or eliminating the presence of intermediaries \cite{sc8494045}.

Since the Ethereum blockchain is "Turing complete"\footnote{\url{https://en.wikipedia.org/wiki/Turing_completeness}} it is possible to create highly customizable smart contracts in many areas, not only financial.

Smart contracts allow trading anything that may have value to the parties involved, for example: money, stocks, property, etc., in a totally transparent manner, without the need for an intermediary and keeping the system free of conflicts of interest \cite{CFOInsights}.
Smart contracts do not only define the rules of the game of any agreement, they are also responsible for the automatic execution of these rules and compliance with the obligations set forth.
In other words, smart contracts are automatically executed lines of code stored on a blockchain that contain predetermined and well-defined rules with no possibility of misunderstanding. 
In this scenario, the space for fraud is reduced and the costs and possible risks of conflicts of interest are also limited \cite{sc8494045,CFOInsights}.
Obviously, as with any other software, it needs to be programmed correctly, free of defects, and in compliance with programming best practices and security guidelines \cite{ethsec} 
\footnote{\url{https://consensys.net/blog/developers/ethereum-smart-contract-security-recommendations/}}
\footnote{\url{https://blog.openzeppelin.com/smart-contract-security-guidelines}}.

\subsection{System definition}

Nowadays, many facilities are faced with the use of hundreds of sheets of paper on which signatures are affixed relative to the informed consent provided by users. 
This information is often archived in a paper format with considerable time and cost, not only in digitizing the documents but also in searching for them. 
In addition, documents that have been signed can be tampered with and/or falsified after the person concerned has signed.
Blockchain and IPFS technology enable the streamlining of the data capture process while offering a very high standard of data quality and security because:
\begin{itemize}
\item	The consent cannot be altered once entered into the blockchain; 
\item	It is impossible to falsify a consent as the insertion can only be done by the holder and any later attempt to alter it is virtually impossible;
\item	An immutable tracking of who accessed the information and at what times is possible;
\item   It is possible to have certain information on the tracking of all the consents given by a person with the relevant timestamp. 
\item	The user has at his disposal a simple and rapid tool to be able to modify the choices made and to be able to renew or remove the consent provided.
\end{itemize}

Since the blockchain, at least in its baseline, is not designed to be able to manage a document repository and/or contain large amounts of data within it, it is, therefore, necessary to use a distributed database that offers high guarantees of security and data integrity. 
For this purpose, the IPFS distributed file system is used to store the template of the informed consent (unfilled informed consent). 

Once the consensus is saved within IPFS, it will be possible to retrieve it via a web address that contains, as a pointer, the hash code of the file itself. In the figure below, you can see an example of how data storage on IPFS works.

\begin{figure}[ht]
\includegraphics[width=9 cm]{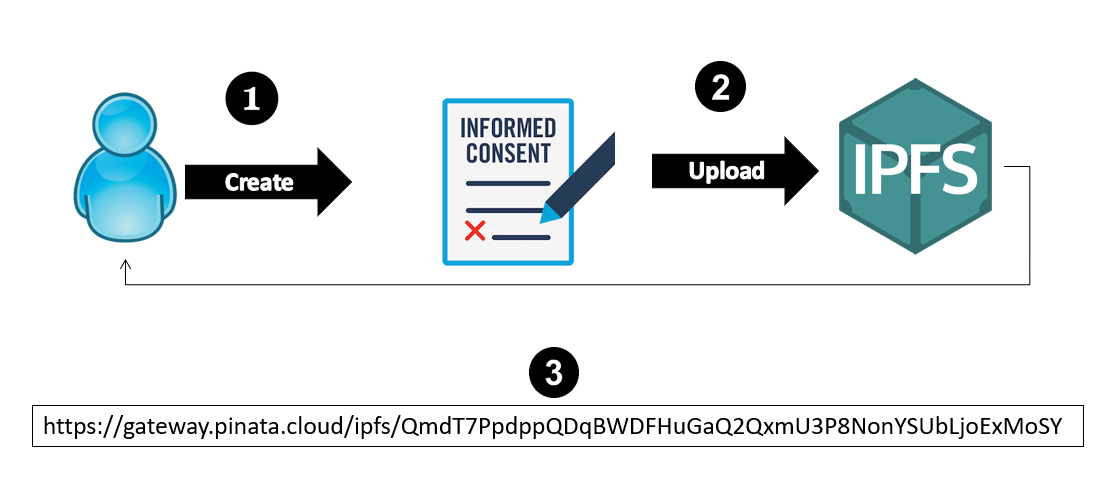}
\caption{File upload in IPFS.\label{fig2}}
\label{fig:fileupoad}
\end{figure}
Once the user has created the document (point 1 - figure \ref{fig:fileupoad}) and saves it within IPFS (point 2 - figure \ref{fig:fileupoad}), the content becomes available via a CID (Content IDentifier) (point 3 - figure \ref{fig:fileupoad}) that corresponds to the hash encoding of the file. \footnote{\url{https://docs.ipfs.io/concepts/how-ipfs-works/\#content-addressing}}

\begin{figure}[ht]
\includegraphics[width=8.8cm]{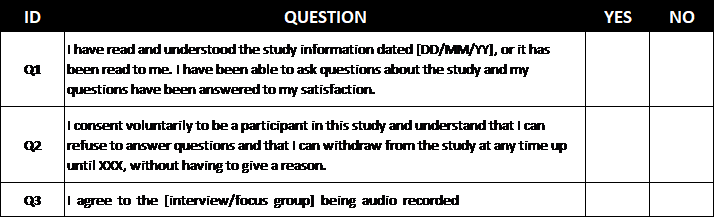}
\caption{Opt-in example\label{fig3}}
\label{fig:optin}
\end{figure}

The file saved to IPFS will be an unfilled informed consent with clearly indicated available options to be selected.
Each field to be selected must be indicated by a letter or number so that it can be easily identified without the possibility of error or confusion, as shown in the following image (Figure \ref{fig:optin}).

\begin{figure}[ht]
\includegraphics[width=9 cm]{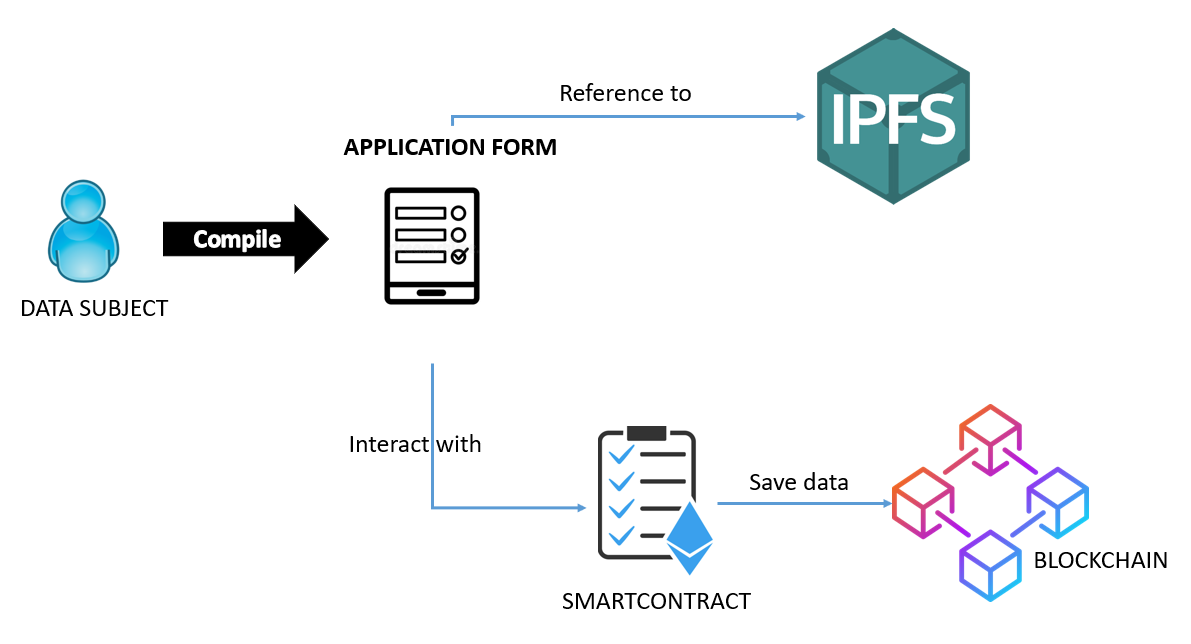}
\caption{Project Schema\label{fig4}}
\label{fig:schema}
\end{figure}

Once the unfilled informed consent  is saved within IPFS, a distributed application (DApp) will propose to the data subject an application form (Figure \ref{fig:schema}) on a web page where he can choose the available options.
The web form will contain the same questions present in the blank informed consent that was saved on IPFS, and the user will be able to choose the same available options (e.g., yes/no).
Once the form is successfully filled out, the application will save the data within the Blockchain as a string in a clear and easily intelligible format. The example transaction array will contain the following data: [Q1 YES, Q2 YES, Q3 NO].

All this will be possible through a smart-contract, specially designed to historicize the data in a suitable format.

The blockchain, once the transaction is validated, will ensure the date and time the data subject entered the information and the choices that were made.
The record saved on the blockchain will not contain any user data except for the wallet that made the data entry.
The wallet is not directly traceable to a user as it is an alphanumeric string in hexadecimal format such as, for example: 0x71C7656EC7ab88b098defB751B7401B5f6d8976F.

To ensure the identity of the data subject performing the transaction, the data controller could use a recognition system similar to the Italian SPID (Digital Public Identity System), or KYC (Know Your Customer, a part of AMLD Anti Money Laundry Directives), storing off the blockchain (off-chain), in a separate, encrypted database, the match between the user and the wallet that authenticated on the DApp and made the input of the consent options.

Off-chain data\cite{ibmoc} are those data that require the ability to be modified or deleted, or are too large to be stored efficiently in the blockchain.
The use of off-chain data storage meets the requirements of the General Data Protection Regulation (GDPR) because it is possible to store sensitive information out of the blockchain so that it can be deleted when needed.
In addition, the use of off-chain systems makes it possible to verify the identity of users by cross-referencing electronic signatures or biometric data with a secure database, sending the data to applications that need to verify the identity of users.

Figure \ref{fig4} summarizes, at a high level, the interactions that occur between the data subject, the IPFS file system and the Ethereum ecosystem (blockchain and smart contract). 
The data subject fills out the web form referring to the blank informed consent saved on IPFS. Then the chosen options are saved on the blockchain via a smart contract. 

\section{Discussion: the Active Informed Consent and the Laws}
\label{sec:system_vs_laws}

The solution proposed in this publication shall allow the data controllers to meet the requirements set out in the data protection law for obtaining valid consent and for processing the data in a data protection compliant manner. 
The following sections include the main principles, which should be considered and complied with, as well as the information on how the tool addresses them or what aspects can be still considered in the further improvement of the technicalities of the tool. 

\subsection{Lawfulness, transparency and fairness (Article 5(1)(a) GDPR)}
The processing of personal data has to always be based on one of the legal bases defined in Article 6 (and in the case of special categories of data – in conjunction with Article 9) of the GDPR. 
The AIC platform allows for collecting consent of the data subjects. This means that in this scenario processing of the data is based on consent as defined in Article 6(1)(a) and 9(2)(a) GDPR and fulfills all the requirements as defined in Article 6(1)(a) GDPR and described in this article (see in particular Table \ref{tab_EC}). 
 
The consent will be provided by the user – a research participant and in the same time a data subject. The user will be able to create a document on a web application, using the template saved on the IPFS. Once the file is saved on IPFS, the DApp will provide the user with an application form on a web page, which will include the aforementioned options/parts to be selected. As soon as the form is successfully filled out, the data will be saved within the blockchain.
Thanks to this solution, the consent will meet the criteria listed in Table \ref{tab_EC}, i.e. will be explicit, freely given, unambiguous and specific, and hence, also informed. 

\begin{itemize}
\item \emph{Explicit} – the consenting person will be provided with all options to agree or disagree and will have the possibility to explicitly indicate his/her choice.
\item \emph{Freely given} – the individual will have a free choice of which point of the data processing he or she agrees to, thanks to the applied opt-in format.
\item \emph{Unambiguous  and specific} - the purpose of each processing operation will be specified in a transparent, clear, and understandable way.
\end{itemize}

All necessary information required by the element of being informed (see Table \ref{tab_EC}) in the consent are fulfilled. 

In this solution, the processing and related aspects are transparently explained at the same time when the data subject receives in the context of the research that she or he would participate in. This information about the processing of the data in a specific research project can be inserted in the IPFS or in another document provided to the user (i.e. data subject) via a web application. The consent filled out by the data subject saved on IPFS will contain all the accompanying information, including the description of a specific study and processing of data. It may be a helpful tool for the data controller, i.e. the entity conducting research and deciding on the means and purposes of data processing in the context of the research, to address the requirement to provide the data subject with appropriate and GDPR-compliant information. This information will also include the information on how consent and data included in the consent form will be processed with the use of the subject AIC tool. Moreover, regarding the blockchain itself, the validation rules may be openly available, and they can be easily communicated to the data subject (i.e. user). \footnote{Luis-Daniel Ibáñez, Kieron O’Hara, and Elena Simperl, On Blockchains and the General Data Protection Regulation, p. 6}
Regarding the transparency of the processing of the data in the AIC system where the data subject will provide his or her consent and the user name (that means the data which allows for clear statement which consent was given by the data subject) the data contained in the blockchain is available to all nodes in the network, and it is also freely available to anyone accessing the network. This feature provides the blockchain with the peculiarity of being transparent, and therefore trust in the system is guaranteed by this feature.

The data included in the consent form is processed fairly - the consent and the data related to the data subject, included in the consent form, is kept in a form as it was given, cannot be altered once entered into the blockchain, and cannot be falsified considering that insertion can be done explicitly by the holder and altering it virtually later is not possible for any other actor.  Moreover, in the case of any unlawful processing that can never be excluded, the blockchain allows for identification of the processor and thanks to that - further prosecution of the committed infringement. \\

\subsection{Purpose limitation (Article 5(1)(b) GDPR)} 

Considering the wording of the purpose limitation principle, data should be collected for specified, explicit and legitimate purposes and not further processed in a manner that is incompatible with those purposes, unless it is processed for selected purposes, such as archiving, histological and scientific research. 

One of the main aims of the AIC tool is to address this basic requirement, by allowing the data subjects to agree to the new purposes of data processing at the time the purpose becomes real and tangible and a new research study and its details are known and can be accurately and transparently described and made understandable to them.  The proposed AIC solution is a remedy to the issue of having the purpose defined too broadly and enables obtaining consent separately for each research study, and therefore allows the data subjects to make their own decision whether they wish to enable certain parties to process their data in a specific context and for a specific purpose.It is due to the fact that within the informed consent template saved on the IPFS platform, it is possible to include this information in great detail and the form and scope required by the GDPR provisions. 

The smart contract allows for including such information about the processing purpose and allows providing consent to specific points in a clear and distinguishable manner, by creating a list that matches the question to the user's answer in a format similar to the following: [Q1 YES, Q2 YES, Q3 NO, ...]. 
Each question (e.g., Q1) is precisely defined and included in the informed consent template and requires a clear answer from the data subject. Each question may also correspond with a different purpose, to which the data subject may or may not consent. The smart contract itself allows or creation of an array of dozens of question/answer pairs in a very simple way. In that manner, the purposes may be added or extended, depending on the future development of the research the data subject participates in and the processing purposes that may appear relevant for certain research study. 

\subsection{Data Minimization (Article 5(1)(c) GDPR), Accuracy (Article 5(1)(d) GDPR and Storage 5(1)(e) GDPR}

The data minimization principle requires that personal data are kept accurate, relevant and limited to what is necessary in relation to the purposes for which they are processed. 

Other two principles which, to some extent, can be addressed by similar means and therefore have been combined in one subsection are “accuracy” and “storage limitation”. 

Accuracy imposes the necessity to keep the data accurate and up to date, and to erase or rectify inaccurate data without delay. In that case, the inaccuracy shall be assessed in relation to the purposes for which the data is processed. 

Storage limitation principle demands keeping personal data in a form, which permits identification of data subjects for no longer than is necessary for the purposes for which the personal data are processed. The only exception, allowing for keeping the data for longer periods, is insofar as the data will be processed exclusively for archiving in public interest, scientific and historical research and statistical purposes, but in that case appropriate security measures imposed by Article 89(1) GDPR have to be adopted. 

Both accuracy and storage limitation principles, requiring rectification or erasure of all inaccurate and not necessary personal data, are closely related to the data subject rights to rectification and erasure. Therefore, the analysis presented below will include the response of the AIC to those rights, hence – indirectly - the response to the principles set out in Article 5(1) GDPR. 

At first sight, the blockchain technology used as part of the AIC system may seem to be a solution that does not meet data minimization, accuracy and storage limitation  requirements, due to its immutability. The data once provided into the chain cannot be altered or changed. What in the situation when a data subject withdraws given consent?  it should be considered that there are means how to address this issue in a sufficient manner, so that it complies with the defined principles, and will indicate how this means has been considered in the AIC. 

As indicated above - with regard to the data minimization principle, in addition to Article 5(1)(b) GDPR, the Recital 39 GDPR  specifies that personal data should be processed only if the purpose of the processing could not reasonably be fulfilled by other means. 

Furthermore, data minimization relates not only to the quantity of data processed, as conventionally understood, but also to the quality of data. From that perspective, it is not necessarily required to erase the data, but rather that the data should not be processed in the shape allowing for identification of a data subject, unless absolutely necessary and that data is pseudonymized or even anonymized whenever possible.\footnote{\url{https://www.europarl.europa.eu/RegData/etudes/STUD/2019/634445/EPRS_STU(2019)634445_EN.pdf}}.  
The purpose of processing of the data on the AIC includes not only the initial research, but also subsequent research possibilities. Therefore, keeping the track of the history of consent changes, especially updating the consent status and maintaining accurate information about data subject's decision on participation in specific research studies and consenting to processing data in that context, justifies the need to process the data. 
Access to the previously added consent “blocks” is important to ensure that the use of data does not go beyond what has been agreed to by the individual, not only in the last “block”, but also in the previously added ones. Additionally to that, the replicated nature of the blockchain distributed databases and the continuous storage of data could be considered to be compliant with purpose limitation.

In the case where data minimization is an effect of the exercise of the right to erase the data, it is true that technically it is impossible to grant the request for erasure made by a data subject when data is registered on a blockchain. Nevertheless, the data could be made inaccessible  thanks to deleting a hash generated by a keyed-hash function or a ciphertext obtained through “state of the art” algorithms and keys and therefore reach an effect being closer to the data erasure. 
With regard to the right to data rectification, also closely related to data minimization and accuracy, it can be argued that even though it is impossible to modify the data once they are in the blockchain, nevertheless, the next transaction can disregard a previously conducted one. 

Following that, a new consent or withdrawal of the consent can cancel the previously made choices or, alternatively, can add to it, if the next transaction is extending the initial consent by including a consent for participation in and for processing of the data in the next research study.
In the AIC, the incomplete data could be modified by a ‘supplementary statement’.  This is possible to be achieved in its distributed ledgers part, as any party who may have access rights, i.e. the data subject or the data controller following the request of the data subject, may add new data to the ledger so that the new information would rectify the previous one. For example, the initially given consent could be modified by adding information about its withdrawal or updated by consent to participate in an additional research initiative.

Additionally, with regard to the data minimization that requires a specific, not a broad consent, AIC addresses the issue of a broad consent perfectly - the data subjects are not required in advance to agree to an unknown number of research studies or types. They can decide which consent to sign based on multiple templates available, and at the same time decide whether they would like to participate in the study and contribute their data thereto at the time the study is defined.

Regarding the compliance with data minimization, accuracy and storage, another right of a data subject should be considered – right to erasure of the data. The response to such request is closely related with appropriate fulfillment of the principles of data minimization, accuracy and storage limitation. A direct deletion of the data from the blockchain can be, however, burdensome and very difficult due to the immutability of the blocks inserted into the chain. Nevertheless, “erasure” does not have to mean “physical destruction” of the data. The GDPR itself does not define the term "erasure" - it may include any kind of obliteration. An alternative solution for physical destruction could be, for e.g.,  anonymization, which would allow for irreversible destruction of the link between the data and the data subject. It should be reminded that the GDPR does not apply to anonymous data. However, in that context it should be taken into account that for “erasure” complete irreversibility is not necessary either. The controller has the right to choose the means of deletion. The important requirement in that regard is to make the data illegible and not available for the controller, and ascertaining that neither controller nor a third party can restore personal identification without disproportionate effort. \footnote{N. Forgo, Z. Skorjanc, Ausgewählte datenschutzrechtliche Fragen im Zusammenhang mit der Personenzertifizierung in der Blockchain, p. 36, accessible here: https://www.wko.at/service/netzwerke/gutachten-personenzertifizierung-blockchain-forgo.pdf}

Following that, there may be ways in the blockchain, which would allow satisfying the right to erasure without achieving the impossible, i.e. deleting the data from the chain. 

It should be also stressed that in the case of AIC, the blockchain is not designed to manage a document repository. For these purposes, the AIC uses IFPS distributed file system. This is where the template of informed consent is stored. Further, this template is made available to the data subject via a web application and the answers provided by the data subject, made based on and according to this template, are stored within the blockchain in a clear and intelligible format. As the outcome of this process, the record stored on the blockchain includes only the information about the wallet that made the data entry, but does not contain any user data, i.e. data of the data subject, which would allow for their direct identification. The wallet itself, as discussed in the previous sections, is not traceable to the user. The whole action is enabled thanks to the application of a smart contract, which allows saving the data in a suitable format.  

Putting personal data “beyond use” could be a solution, which would allow achieving data minimization and data accuracy would be appropriate response to the right to erasure.
\footnote{\url{https://ico.org.uk/media/for-organisations/documents/1475/deleting_personal_data.pdf}, p. 4}  
“Beyond use” could be understood as just moving the data out of a “live” system, as deleting them is not possible due to technical reasons, as such deletion would involve deletion of other information held in the same batch. It is what has been considered and applied in AIC -  whenever there is no intention to access or process the data and keep it “live”, the data is put off-chain, in a separate, encrypted database \footnote{Ibid, p. 8, M. Finck, “Blockchains and Data Protection in the European Union,” SSRN Electronic Journal, 2017; }. In the case where the deletion of the data is required, the record on the off-chain database which links the data subject's name to the wallet that performed the transaction on the blockchain can be erased. In that way, the direct link between the data subject and the choices made will be deleted and the identification of the data subject not possible.  
On top of it, once the data subject requests for erasure, the data off-chain, which could allow to identify the data subject, can be erased, as the data stored off-chain may be a subject to modification or deletion (the system here is not immutable). The same response could be used in the context of the right to rectification - the incorrect record can be erased and new, correct ones, can be added to the chain as a new transaction. The first one would be still in the blockchain, but leading to no data allowing for identification of a data subject.

In that manner the AIC, build on IPFS, blockchain and smart contracts, responds to the discussed the GDPR principles, finding the balance between the data protection requirements and the usefulness of the solution. 

\subsection{Confidentiality and integrity (Article 5(1)(f) GDPR)}

This principle requires that personal data should be processed in a manner ensuring security ‘including protection against unauthorized or unlawful processing and against accidental loss, destruction or damage, using appropriate technical or organizational measures (TOMs)’. In the context of blockchain and IPFS technology, included in the AIC system – the solution, providing pseudonymisation and encryption, responds to the confidentiality and integrity requirements, as it allows hiding the identity of the user. Additionally, the data access and data modification rights are very limited and are granted only to the user, and the access to the information and the time when it has been done is tracked. It means that only the data subject can make a transaction on the blockchain with his or her own private key. 
The distributed system is also highly resistant to failure or cyber-attacks \cite{1084633746,Storm2012} - the usage of Ethereum blockchain, which is a distributed system, gives the entire network high availability and reliability, and the fact that each node has a copy of the entire database ensures that there is no single point of failure, as any damage to a peer in the network does not affect the overall operation of the system. 
The usage of smart contracts, when programmed correctly and in line with security guidelines and programming best practices, also allows for reducing the space for fraud.
Moreover, keeping the data on an off-chain transactional data storage, as described in the section "System definition", is an additional aspect, which would support compliance with the data protection requirements related to the application of TOMs.

\section{Conclusions}
\label{sec:conclusions}

Machine learning (ML) applied to precision medicine is on a cliff edge: if it does not learn to fly, it will deeply fall down.
ML needs data to find regularities. Yet, in medicine, data represent sensitive information about patients, which cannot be used without their consent. 

In this paper, we presented a possibility to let Machine Learning better contribute to the health system. The proposed Active Informed Consent (AIC) allows machine learning to legally use personal data when learning, and it puts patients at the center of their privacy management. Patients are active subjects, responsible for how their data is used. We carefully analyzed the compliance of a technological tool to the legal intricacies of protecting the privacy of European Citizens.

We suggest using this technical solution in hospitals and health systems around Europe in order to foster better research in precision medicine for public health, since it will unlock the possibility of reusing data in many different projects.


\bibliographystyle{plain}

\bibliography{bibliography.bib}

\begin{thebibliography}{10}

\bibitem{ethsec}
{ Ethereum Foundation}.
\newblock Smart contract security guidelines, 2020.

\bibitem{art/transp}
Cc~Agbo and Qusay Mahmoud.
\newblock Blockchain in healthcare: Opportunities, challenges, and possible
  solutions.
\newblock {\em International Journal of Healthcare Information Systems and
  Informatics}, 15:82--97, 07 2020.

\bibitem{1152290190}
JAI~SING. ARUN.
\newblock {\em BLOCKCHAIN FOR BUSINESS.}
\newblock NIELSEN BOOKDATA, [S.l.], 2019.

\bibitem{ibmoc}
Michael Ault.
\newblock Why new off-chain storage is required for blockchains document
  version 1.0, 10 2018.

\bibitem{babar_evaluating_2021:image_analysis}
Zaheer Babar, Twan van Laarhoven, Fabio~Massimo Zanzotto, and Elena Marchiori.
\newblock Evaluating diagnostic content of {AI}-generated radiology reports of
  chest {X}-rays.
\newblock {\em Artificial Intelligence in Medicine}, 116:102075, June 2021.

\bibitem{1032154406}
Imran Bashir.
\newblock Mastering blockchain : Distributed ledger technology,
  decentralization, and smart contracts explained, 2nd edition, 2018.

\bibitem{Budin-Ljosne2017}
Isabelle Budin-Lj{\o}sne, Harriet J.~A. Teare, Jane Kaye, Stephan Beck,
  Heidi~Beate Bentzen, Luciana Caenazzo, Clive Collett, Flavio D'Abramo, Heike
  Felzmann, Teresa Finlay, Muhammad~Kassim Javaid, Erica Jones, Vi{\v{s}}nja
  Kati{\'{c}}, Amy Simpson, and Deborah Mascalzoni.
\newblock Dynamic consent: a potential solution to some of the challenges of
  modern biomedical research.
\newblock {\em BMC Medical Ethics}, 18(1):4, Jan 2017.

\bibitem{budin-ljosne_dynamic_2017}
Isabelle Budin-Ljøsne, Harriet J.~A. Teare, Jane Kaye, Stephan Beck,
  Heidi~Beate Bentzen, Luciana Caenazzo, Clive Collett, Flavio D’Abramo,
  Heike Felzmann, Teresa Finlay, Muhammad~Kassim Javaid, Erica Jones, Višnja
  Katić, Amy Simpson, and Deborah Mascalzoni.
\newblock Dynamic {Consent}: a potential solution to some of the challenges of
  modern biomedical research.
\newblock {\em BMC Medical Ethics}, 18(1):4, January 2017.

\bibitem{buterin2009next}
V~Buterin.
\newblock A next generation smart contract and decentralized application
  platform, (january), 2009.

\bibitem{chen_machine_2017:prediction}
Jonathan~H. Chen and Steven~M. Asch.
\newblock Machine {Learning} and {Prediction} in {Medicine} — {Beyond} the
  {Peak} of {Inflated} {Expectations}.
\newblock {\em The New England journal of medicine}, 376(26):2507--2509, June
  2017.

\bibitem{dara_machine_2022:drug_discovery}
Suresh Dara, Swetha Dhamercherla, Surender~Singh Jadav, Ch~Madhu Babu, and
  Mohamed~Jawed Ahsan.
\newblock Machine {Learning} in {Drug} {Discovery}: {A} {Review}.
\newblock {\em Artificial Intelligence Review}, 55(3):1947--1999, 2022.

\bibitem{CFOInsights}
{Deloitte Development LLC}.
\newblock Getting smart about smart contracts, 2016.

\bibitem{7a6a23e4a9864a0e80f47484b1bfe2cf}
Johan Hansen, Petra Wilson, Eline Verhoeven, M.~Kroneman, Mary Kirwan, Robert
  Verheij, and E-B. {van Veen}.
\newblock {\em Assessment of the EU Member States' rules on health data in the
  light of GDPR}.
\newblock European Union, 2021.

\bibitem{Kaye2015}
Jane Kaye, Edgar~A. Whitley, David Lund, Michael Morrison, Harriet Teare, and
  Karen Melham.
\newblock Dynamic consent: a patient interface for twenty-first century
  research networks.
\newblock {\em European Journal of Human Genetics}, 23(2):141--146, Feb 2015.

\bibitem{bcipfs}
Shivansh Kumar, Aman~Kumar Bharti, and Ruhul Amin.
\newblock Decentralized secure storage of medical records using blockchain and
  ipfs: A comparative analysis with future directions.
\newblock {\em SECURITY AND PRIVACY}, 4(5):e162, 2021.

\bibitem{10.1093/jamia/ocx068}
Tsung-Ting Kuo, Hyeon-Eui Kim, and Lucila Ohno-Machado.
\newblock {Blockchain distributed ledger technologies for biomedical and health
  care applications}.
\newblock {\em Journal of the American Medical Informatics Association},
  24(6):1211--1220, 09 2017.

\bibitem{1084633746}
Dac-Nhuong Le, Raghvendra Kumar, Kishore Mishra, Brojo, Manju Khari, and Moy
  Chatterjee, Jyotir.
\newblock Cyber security in parallel and distributed computing : concepts,
  techniques, applications and case studies, 2019.

\bibitem{maier_gentle_2019:image_analysis}
Andreas Maier, Christopher Syben, Tobias Lasser, and Christian Riess.
\newblock A gentle introduction to deep learning in medical image processing.
\newblock {\em Zeitschrift Fur Medizinische Physik}, 29(2):86--101, May 2019.

\bibitem{sc8494045}
Bhabendu~Kumar Mohanta, Soumyashree~S Panda, and Debasish Jena.
\newblock An overview of smart contract and use cases in blockchain technology.
\newblock In {\em 2018 9th International Conference on Computing, Communication
  and Networking Technologies (ICCCNT)}, pages 1--4, 2018.

\bibitem{pmid31729900}
M.~Prictor, M.~A. Lewis, A.~J. Newson, M.~Haas, S.~Baba, H.~Kim, M.~Kokado,
  J.~Minari, F.~Molnár-Gábor, B.~Yamamoto, J.~Kaye, and H.~J.~A. Teare.
\newblock {{D}ynamic {C}onsent: {A}n {E}valuation and {R}eporting {F}ramework}.
\newblock {\em J Empir Res Hum Res Ethics}, 15(3):175--186, 07 2020.

\bibitem{8287055}
Deepak Puthal, Nisha Malik, Saraju~P. Mohanty, Elias Kougianos, and Chi Yang.
\newblock The blockchain as a decentralized security framework [future
  directions].
\newblock {\em IEEE Consumer Electronics Magazine}, 7(2):18--21, 2018.

\bibitem{Storm2012}
Christian Storm.
\newblock {\em Fault Tolerance in Distributed Computing}, pages 13--79.
\newblock Vieweg+Teubner Verlag, Wiesbaden, 2012.

\bibitem{9045940}
Jin Sun, Xiaomin Yao, Shangping Wang, and Ying Wu.
\newblock Blockchain-based secure storage and access scheme for electronic
  medical records in ipfs.
\newblock {\em IEEE Access}, 8:59389--59401, 2020.

\bibitem{Szabo_1997}
Nick Szabo.
\newblock Formalizing and securing relationships on public networks.
\newblock {\em First Monday}, 2(9), Sep. 1997.

\bibitem{Teare2021}
Harriet J.~A. Teare, Megan Prictor, and Jane Kaye.
\newblock Reflections on dynamic consent in biomedical research: the story so
  far.
\newblock {\em European Journal of Human Genetics}, 29(4):649--656, Apr 2021.

\bibitem{vamathevan_applications_2019:drug_discovery}
Jessica Vamathevan, Dominic Clark, Paul Czodrowski, Ian Dunham, Edgardo Ferran,
  George Lee, Bin Li, Anant Madabhushi, Parantu Shah, Michaela Spitzer, and
  Shanrong Zhao.
\newblock Applications of machine learning in drug discovery and development.
\newblock {\em Nature Reviews Drug Discovery}, 18(6):463--477, June 2019.

\bibitem{Yaqoob2022}
Ibrar Yaqoob, Khaled Salah, Raja Jayaraman, and Yousof Al-Hammadi.
\newblock Blockchain for healthcare data management: opportunities, challenges,
  and future recommendations.
\newblock {\em Neural Computing and Applications}, 34(14):11475--11490, Jul
  2022.

\bibitem{Zanzotto2019243}
F.M. Zanzotto.
\newblock Viewpoint: Human-in-the-loop artificial intelligence.
\newblock {\em Journal of Artificial Intelligence Research}, 2019.

\bibitem{zhang_machine_2017:drug_discovery}
Lu~Zhang, Jianjun Tan, Dan Han, and Hao Zhu.
\newblock From machine learning to deep learning: progress in machine
  intelligence for rational drug discovery.
\newblock {\em Drug Discovery Today}, 22(11):1680--1685, November 2017.

\bibitem{zhang_radiological_2019:image_analysis}
Zhenwei Zhang and Ervin Sejdić.
\newblock Radiological images and machine learning: trends, perspectives, and
  prospects.
\newblock {\em Computers in biology and medicine}, 108:354--370, May 2019.

\bibitem{bttech}
Zibin Zheng, Shaoan Xie, Hong-Ning Dai, Xiangping Chen, and Huaimin Wang.
\newblock An overview of blockchain technology: Architecture, consensus, and
  future trends.
\newblock 06 2017.

\end{thebibliography}

\end{document}